\documentclass{emulateapj}
\usepackage{apjfonts}
\journalinfo{The Astrophysical Journal, 691, 2009 January 20, in press}

\shorttitle{ION TEMPERATURES IN THE SOLAR CORONA}
\shortauthors{LANDI \& CRANMER}

\newcommand{\etal}{{et al.\ }}
\def\ion[#1 #2]{#1\,{\sc #2}}
\def\lamb[#1]{#1\,{\AA}}
\def\lambr[#1-#2]{{{#1}--{#2}\,{\AA}}}

\begin{document}

\title{Ion Temperatures in the Low Solar Corona: Polar Coronal Holes
at Solar Minimum}

\author{E. Landi}
\affil{Artep, Inc. at Naval Research Laboratory, 4555 Overlook Ave.
S.W., 20375-5320, Washington DC}
\author{S. R. Cranmer}
\affil{Harvard-Smithsonian Center for Astrophysics,
60 Garden Street, Cambridge, MA 02138}

\begin{abstract}
In the present work we use a deep-exposure spectrum taken by the SUMER 
spectrometer in a polar coronal hole in 1996 to measure the ion temperatures
of a large number of ions at many different heights above the limb between
0.03 and 0.17 solar radii.
We find that the measured ion temperatures are almost always
larger than the electron temperatures and exhibit a
non-monotonic dependence on the charge-to-mass ratio.
We use these measurements to provide empirical constraints to a
theoretical model of ion heating and acceleration based on gradually
replenished ion-cyclotron waves.
We compare the wave power required to heat the ions to the
observed levels to a prediction based on a model of anisotropic
magnetohydrodynamic turbulence.
We find that the empirical heating model and the turbulent cascade
model agree with one another, and explain the measured ion temperatures,
for charge-to-mass ratios smaller than about 0.25.
However, ions with charge-to-mass ratios exceeding 0.25
disagree with the model; the wave power they require to be heated
to the measured ion temperatures shows an increase with charge-to-mass
ratio (i.e., with increasing frequency) that cannot be explained by a
traditional cascade model.  We discuss possible additional processes
that might be responsible for the inferred surplus of wave power.
\end{abstract}


\section{Introduction}

In order to understand the physical processes that heat the solar
corona and accelerate the solar wind, theories must be able to
predict measurements of the plasma parameters in the regions that
are being heated and accelerated.
In the low-density, open-field regions that correspond to polar
coronal holes on the Sun, the plasma is expected to become
collisionless very near the solar surface, and thus the individual
particle species (e.g., protons, electrons, and heavy ions) may
exhibit diverging properties.
These differences are key probes of the microscopic kinetic physics
that drives the heating and acceleration.
The Solar Ultraviolet Measurements of Emitted Radiation
(SUMER) spectrometer aboard the {\em Solar and Heliospheric
Observatory} ({\em SOHO}) has measured these properties for more
than a decade (see, e.g., Wilhelm et al.\  1995, 1997).

Inspired by both the SUMER observations presented below and other
remote-sensing and in~situ measurements, we focus on the damping of
ion cyclotron resonant Alfv\'{e}n waves as a natural means of
preferentially heating and accelerating coronal ions.
Efficient kinetic interactions between ion cyclotron waves and particle
velocity distributions in the solar wind have been studied for several
decades (e.g., Abraham-Shrauner \& Feldman 1977; Hollweg \& Turner 1978;
Marsch \etal  1982; Hollweg 1986; Tu \& Marsch 1997; Cranmer \etal 1999a;
Hollweg \& Isenberg 2002; Cranmer 2002; Kohl \etal  2006). Despite much
intense theoretical work, the origin of the high-frequency ($10^2$--$10^4$~Hz)
cyclotron waves in the solar corona is not yet known. Most ideas for the
generation of these waves fall into two classes: (1) excitation at the
base of the corona, and (2) gradual growth or replenishment of cyclotron
waves throughout the corona and solar wind. The first idea has been
proposed to occur from impulsive processes taking place in magnetically
complex regions (e.g., reconnection in microflares; see Axford \&
McKenzie 1992; Ruzmaikin \& Berger 1998). The second scenario has
been proposed to occur mainly from magnetohydrodynamic (MHD) turbulent
cascade (Isenberg \& Hollweg 1983; Cranmer \& van Ballegooijen 2003;
Chandran 2005), but other processes such as kinetic microinstabilities
have also been suggested. Indirect support for the second idea of
``gradual replenishment'' has come from empirical studies of radio
scintillations (Hollweg 2000) and theoretical models of the charge
and mass dependence of the wave damping (Cranmer 2000, 2001). In our
paper, we will base our model on the gradual replenishment scenario.

One of the main predictions of ion-cyclotron models is preferential heating 
and acceleration of ions in the solar wind, with the effect of waves on ions
being dependent on the $Z_{i}/A_{i}$ ratio, where $Z_{i}$ is the
ion's charge and $A_{i}$ its atomic mass (with both usually specified
in proton units).
One of the observables that can best constrain theoretical models
of ion-cyclotron waves is the behavior of the {\em ion temperature}
(i.e., the magnitude of ion random motions on microscopic scales).
These models are also dependent on the overall amplitude of MHD
waves in the corona, which can also be constrained empirically by
the so-called {\em non-thermal} component of emission line profiles.

Measuring ion temperatures and non-thermal velocities is however a very
difficult task, as these are two separate physical quantities that have
to be determined from one single observable, the line width.
Thus, some assumptions have to be made.
However, these limit the accuracy of the measurements.
In the past, most assumptions were taken on the value of $T_i$,
since the quantity of interest was the non-thermal velocity $v_{\rm nth}$.
The most common assumptions were $T_i=T_e$, or $T_i=T_M$, where $T_M$
is the temperature of formation of the line, usually taken as the
temperature of maximum abundance of 
the ion emitting the line. A more elaborate
set of assumptions were adopted by Seely \etal (1997),
who assumed that ions formed in the same temperature range have the
same $T_i$; this allowed them to determine both
$T_i$ and $v_{\rm nth}$ from SUMER/{\em{SOHO}} observations.
Tu \etal (1998), on the contrary,
assumed that $v_{\rm nth}$ is the same for all ions.
This allowed them to set an upper limit to $v_{nth}$, and
conservatively assumed $v_{\rm nth}=0$ as a lower limit.
From these limits, the range of possible $T_i$ values could be determined.
In doing this, Tu \etal (1998) limited the number of assumptions to the
minimum and chose not to determine $v_{\rm nth}$,
so that they could set rather precise limits to $T_i$ of the ions 
in their dataset.
Dolla \& Solomon (2008) went a step further by making assumptions 
not on the values of $T_i$ or $v_{\rm nth}$, but on the physical
process that affects them.
They assumed that: (1) non-thermal velocities are due to
undamped Alfv\'{e}n waves, and (2) that \ion[Mg x] was not
heated by them so that its $T_i$ was constant and
could be used to measure $v_{\rm nth}$.

The aim of the present work is twofold.
First, we will extend the analysis done
by Tu \etal (1998) by applying their method to spectra emitted by
coronal hole plasmas at the north pole, that include many more lines
from many more ions than considered by them.
We will repeat the $T_i$ measurements for as many 
heights above the solar limb as made possible by the
signal-to-noise ratio and instrument scattered light.
This has been achieved by using deep-exposure SUMER 
spectra encompassing the entire wavelength range (500--1500~\AA)
included by detector~B on SUMER.

Second, we will use the measured ion temperature values as constraints for
semi-empirical theoretical models of ion heating and acceleration 
based on the damping of ion cyclotron resonant Alfv\'{e}n waves. We will 
develop a model that is mainly intended to be illustrative of the proper 
ranges of plasma parameters that are consistent with the observations in 
the low corona, but not yet self-consistent enough to be considered a
definitive or unique explanation of the SUMER data set.\footnote{%
Specifically, these models do {\em not} try to validate or disprove the
assumption that the ion cyclotron waves are created by an anisotropic
turbulent cascade; they only try to work out the implications of
such an assumption.}

The first part of the paper deals with the ion temperature measurements.
Section~\ref{sec3} introduces the method of analysis and Section~\ref{sec2} 
describes the observations studied in the present work. Our measurements
are reported in Section~\ref{sec4}. The second part of the paper describes
the comparison with the theoretical model. In Section~\ref{sec5} the 
separation of line widths into thermal and non-thermal parts is discussed 
in more detail. The physical processes assumed 
in the semi-empirical models are given in Section~\ref{sec6}, and a 
comparison with a purely theoretical model of MHD
turbulence is made in Section~\ref{sec7}.
The model results are shown and described in Section~\ref{sec8}.
Section~\ref{conclusions} summarizes the present work.

\section{Method of Analysis}
\label{sec3}

In this work, we have applied the analysis method described by Tu \etal (1998),
that we briefly summarize here.
The measured full width half maximum (FWHM) $d\lambda$
of a spectral line can be related to the dynamical status of the
emitting ion through the equation
\begin{equation}
d\lambda={{\lambda_0}\over{c}}\sqrt{4\ln 2{\left({\frac{2 k_{\rm B} T_i}{M_i}+v_{\rm nth}^2}\right)}}
\label{fwhm_1}
\end{equation}
where $d\lambda$ is the FWHM of the line after the instrumental
broadening has been removed, $T_i$ and $M_{i} = m_{p} A_{i}$ ($m_p$ is
the proton mass) are the ion's temperature and mass, $k_{\rm B}$
is Boltzmann's constant, $c$ is the speed of light, $\lambda_0$ is
the rest wavelength of the line, and $v_{\rm nth}$ is the ion's
non-thermal velocity.
From equation~(\ref{fwhm_1}),
the ion temperature given by each line can not be larger
than the value obtained when $v_{\rm nth}=0$:
\begin{eqnarray}
T_i & \leq & T_{{\rm max,}i}  \nonumber \\
T_{{\rm max,}i} & = & {{m_pc^2}\over{8 k_{\rm B} \ln 2}}A_i
 {\left({{{d\lambda}\over{\lambda_0}}}\right)}_i^2 =
   1.96\times 10^{12} A_{i}
   {\left({{{d\lambda}\over{\lambda_0}}}\right)}_i^2
   \,\, \mbox{K}
\label{fwhm_2}
\end{eqnarray}
where $A_i$ is the atomic mass.
In the same way, the non-thermal 
velocity can not be larger than the value obtained when $T_i=0$:
\begin{eqnarray}
v_{\rm nth} & \leq & v_{{\rm max,}i}  \nonumber \\
v_{{\rm max,}i} & = & \sqrt{{{c^2}\over{4\ln 2}}{{\left({{{d\lambda}\over{\lambda_0}}}\right)}_i^2}}
        = 1.80\times 10^{10} {\left({{{d\lambda}\over{\lambda_0}}}\right)}_{i}
  \,\, \mbox{cm} \,\, \mbox{s}^{-1} .
\label{fwhm_3}
\end{eqnarray}
In order to determine a lower limit to $T_i$,
it is necessary to make an assumption on $v_{\rm nth}$.
The method devised by Tu \etal (1998) allows us to determine the minimum
$T_i$ values compatible with a set of measured FWHM values by assuming
that $v_{\rm nth}$ is the same for all ions.
If we define $v_{\rm min}$ as the smallest among 
the $v_{{\rm max,}i}$ values obtained from a set of lines, the ion temperature
of each ion must be larger than the value obtained when $v_{\rm nth}=v_{\rm min}$.
If we indicate with $j$ the line for which $v_{\rm nth}=v_{\rm min}$,
the range of ion temperatures compatible with the 
FWHM measured with line $i$ is given by
\begin{equation}
{{m_pc^2}\over{8 k_{\rm B} \ln 2}}A_i
{\left[{{\left({{{d\lambda}\over{\lambda_0}}}\right)}_i^2-{\left({{{d\lambda}\over{\lambda_0}}}\right)}_j^2}\right]} 
\leq T_i \leq {{m_pc^2}\over{8 k_{\rm B} \ln 2}}A_i{\left({{{d\lambda}\over{\lambda_0}}}\right)}_i^2 
\label{fwhm_4}
\end{equation}
Note that the lower limit of $T_i$ for line $j$ is zero. The interval defined
by equation~(\ref{fwhm_4})
is not to be considered the ``error bar'' due to experimental uncertainties in 
the measurement of the line's FWHM; it indicates the range of equally possible 
$T_i$ values allowed by the observed FWHM value. Uncertainties in the measured 
$d\lambda$ are propagated to the limit values of $T_i$ in
equation~(\ref{fwhm_4}), 
and widen the range of possible ion temperatures. 
This method of analysis has the advantage of reducing the assumptions on $T_i$
and $v_{\rm nth}$ to the very minimum. However, it only provides us with an upper
limit to $v_{\rm nth}$.

\section{Observations and Data Reduction}
\label{sec2}

\subsection{Observations}

\begin{figure}
\epsscale{1.1}
\plotone{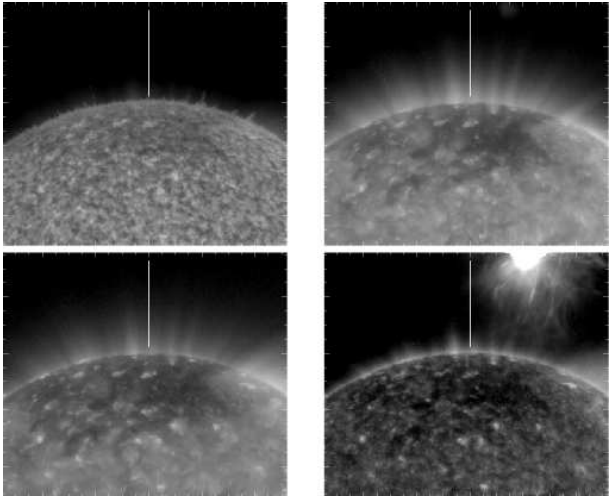}
\caption{\label{eit} Field of view of the SUMER slit, superimposed to
EIT images in all four filters: {\em Top left:} \ion[He ii] ($5\times 10^4$~K);
{\em Top right:} \ion[Fe ix-x] ($6.3\times 10^5$~K); {\em Bottom left:}
\ion[Fe xii] ($1.4\times 10^6$~K); {\em Bottom right:} \ion[Fe xv]
($2.2\times 10^6$~K). The EIT images were taken on the same day as
SUMER (3 November 1996) between 1~UT and 2~UT.}
\end{figure}

Observations were taken on 1996 November 3, and consist of a full spectral
scan encompassing the entire SUMER wavelength range carried out over a
polar coronal hole. The center of the
slit field of view was (0\arcsec,1149\arcsec),
so that coronal hole plasma was observed at heliocentric distances
between 1.03 solar radii ($R_{\odot}$) and 1.33 $R_{\odot}$.
The instrument configuration allows the recording of the solar spectrum
in $\simeq 43$~\AA-wide subsections of the 500--1600~\AA\ spectral range;
in the present observation the full SUMER range was observed in 61
subsections, each shifted from the previous one by 
$\simeq 13$~\AA.
This ensured that all lines were observed on the more sensitive KBr-coated
portion of the detector at least once.
Each spectral subsection was observed with SUMER
detector~B with 1200~s exposure time,
so that the final effective wavelength range of the 
dataset we studied is 500--1500~\AA.
An image of the SUMER field of view superimposed to 
EIT images is shown in Figure~\ref{eit}.

\subsection{Data reduction and scattered light}

\begin{deluxetable}{lrcrcc}
\tablewidth{0pt}
\tabletypesize{\normalsize}
\tablecaption{\label{lines} Lines used for the present study.
$R_{\rm max}$ (in units of solar radii from disk center)
indicates the maximum distance beyond which the line could not be used
due to scattered light contamination.}
\tablehead{\colhead{Ion} & \colhead{Wvl. (\AA)} & \colhead{$R_{\rm max}$}
& \colhead{Seq.} & \colhead{$\log T_{\rm max}$} & \colhead{$Z_{i}/A_{i}$}}
\tablecolumns{6}
\startdata
\ion[Na ix]     &  681.72 & 1.171 & Li & 5.92 & 0.348 \\
\ion[Si ix]     &  694.90 & 1.151 &  C & 6.05 & 0.285 \\
\ion[Ar viii]   &  700.25 & 1.100 & Na & 5.61 & 0.175 \\
\ion[Mg ix]     &  706.06 & 1.171 & Be & 5.99 & 0.329 \\
\ion[Fe viii]   &  721.26 & 1.049 &  K & 5.88 & 0.125 \\
\ion[Mg ix]     &  749.55 & 1.171 & Be & 5.99 & 0.329 \\
\ion[Mg viii]   &  762.65 & 1.059 &  B & 5.91 & 0.288 \\
\ion[Mg viii]   &  769.38 & 1.079 &  B & 5.91 & 0.288 \\
\ion[Ne viii]   &  770.42 & 1.171 & Li & 5.80 & 0.347 \\
\ion[Mg viii]   &  772.29 & 1.090 &  B & 5.91 & 0.288 \\
\ion[Ne viii]   &  780.34 & 1.171 & Li & 5.80 & 0.347 \\
\ion[Mg viii]   &  782.37 & 1.151 &  B & 5.91 & 0.288 \\
\ion[Ca ix]     &  821.22 & 1.049 & Mg & 5.80 & 0.200 \\
\ion[Mg vii]    &  854.64 & 1.079 &  C & 5.80 & 0.247 \\
\ion[Mg vii]    &  868.08 & 1.110 &  C & 5.80 & 0.247 \\
\ion[Ne vii]    &  887.24 & 1.028 & Be & 5.71 & 0.297 \\
\ion[Ne vii]    &  895.16 & 1.100 & Be & 5.71 & 0.297 \\
\ion[S vi]      &  933.41 & 1.079 & Na & 5.38 & 0.156 \\
\ion[Ne vii]    &  973.33 & 1.028 & Be & 5.71 & 0.297 \\
\ion[Fe x]      & 1028.04 & 1.161 & Cl & 6.08 & 0.161 \\
\ion[O vi]      & 1031.93 & 1.171 & Li & 5.58 & 0.313 \\
\ion[O vi]      & 1037.63 & 1.171 & Li & 5.58 & 0.313 \\
\ion[Si vii]    & 1049.26 & 1.079 &  O & 5.79 & 0.214 \\
\ion[Al vii]    & 1053.88 & 1.059 &  N & 5.79 & 0.222 \\
\ion[Al vii]    & 1056.79 & 1.059 &  N & 5.79 & 0.222 \\
\ion[Al viii]   & 1057.86 & 1.069 &  C & 5.94 & 0.259 \\
\ion[Ca x]      &  557.76 & 1.110 & Na & 5.87 & 0.225 \\
\ion[Ne vi]     &  558.59 & 1.028 &  B & 5.63 & 0.248 \\
\ion[Ne vi]     &  562.80 & 1.028 &  B & 5.63 & 0.248 \\
\ion[Ca x]      &  574.01 & 1.110 & Na & 5.87 & 0.225 \\
\ion[Mg x]      &  609.80 & 1.171 & Li & 6.04 & 0.370 \\
\ion[N v]       & 1238.82 & 1.079 & Li & 5.24 & 0.286 \\
\ion[Fe xii]    & 1242.00 & 1.141 &  P & 6.22 & 0.197 \\
\ion[N v]       & 1242.81 & 1.069 & Li & 5.24 & 0.286 \\
\ion[Mg x]      &  624.97 & 1.161 & Li & 6.04 & 0.370 \\
\ion[Si viii]   & 1440.51 & 1.028 &  N & 5.92 & 0.249 \\
\ion[Si viii]   & 1445.76 & 1.161 &  N & 5.92 & 0.249 \\
\ion[Fe xi]     & 1467.07 & 1.100 &  S & 6.14 & 0.179 \\
\enddata
\end{deluxetable}

Raw data were reversed and corrected from flat field and geometrical
distortions using the standard SUMER software available in the
SolarSoft package for IDL.  Corrections for deadtime 
and local gain depression were not necessary because count rates were always 
sufficiently low. A large set of lines was selected in order to sample as wide
a range of $Z_{i}/A_{i}$ ratios as possible with bright,
isolated lines that minimized uncertainties tied to line profile fitting.
These lines are listed in Table~\ref{lines}.
In order to increase the signal-to-noise, the SUMER slit was
divided into 10 pixel-wide bins along the slit direction: within 
each bin, the emission was averaged.
The profile of each line was fitted with a 
Gaussian profile, obtaining line centroid, width and total intensity.

\begin{figure}
\includegraphics[width=6.4cm,angle=90]{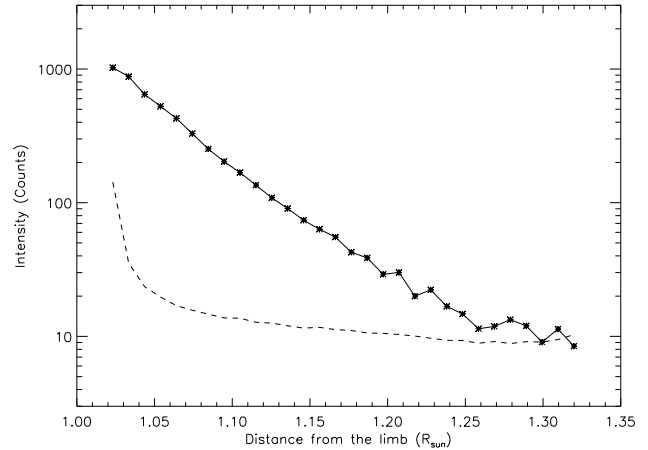}
\caption{\label{scatt} Intensity of the \ion[Mg ix] 706.06~\AA\ line
as a function of distance from the limb (in counts). The scattered light
intensity is superimposed as a dashed line, and it has been normalized
to match the \ion[Mg ix] intensity averaged over the three positions
farthest from the solar limb.}
\end{figure}

In the present work, we are primarily interested in line widths, but line 
intensities were used to determine the amount of scattered light contamination. 
Instrument-scattered light consists of disk radiation scattered by mirror 
micro-roughness, that forms on the detector a ghost, unshifted spectrum of 
the solar disk that contaminates true local emission. A significant scattered 
light blending may alter line profiles in ways that are not easy to assess 
and correct. Before attempting an analysis of line widths, it is important 
to check that the scattered light contribution is negligible and, if not 
negligible, to remove it.
Scattered light can be distinguished from true coronal emission by considering 
the variations of the measured intensity with distance from the limb, because 
true emission decreases much faster than instrument-scattered light. In order 
to evaluate the amount of scattered light contaminating each line, we first 
determined the rate of decrease of scattered light intensity with distance 
from the limb. This was done first by measuring the normalized intensity 
versus height relationship (henceforth {\em scattered light profile}) of a 
few chromospheric and transition region lines which, outside the solar disk, 
are almost entirely made of scattered light. We then assumed that the 
emission of the selected coronal lines measured at the farthest distance 
from the limb is entirely composed of scattered light, and used the scattered 
light profile to evaluate the absolute intensity of instrument-scattered
emission for each coronal line at all heights.
An example of comparison between scattered 
light and true coronal emission is given in Figure~\ref{scatt}
for the \ion[Mg ix] line at 706.06~\AA. It is important to note that 
this method actually overestimates the scattered light emission. We considered 
the scattered light to be negligible when its intensity, as estimated using 
this method, is 25\% or less of the measured intensity of a coronal line. 

This method allowed us to determine the maximum height
$R_{\rm max}$ below which the scattered light can be safely considered
negligible.
For the strongest lines, we could stretch our measurements
up to 1.17 $R_{\odot}$, while only 
the closest bin provided intensity for the weakest lines.
Table~\ref{lines} also reports the value of $R_{\rm max}$ for each line.

The presence of scattered light in line widths can also be checked
independently by considering the quantity $d\lambda / \lambda_0$.
In fact, from equation~\ref{fwhm_1} it is easy to see that
$d\lambda / \lambda_0$
only depends on physical properties of the ion and so 
it is the same for all lines of the same ion, unless some problem
affects the profile of a line. We calculated this quantity for all the 
ions in Table~\ref{lines} with more than one line in order to check 
whether there was agreement at all heights up to $R_{\rm max}$. In the 
few cases where $d\lambda/\lambda_0$ was not constant within the same 
ion, we have investigated the cause of the disagreement, usually due 
to poor signal-to-noise or to blending lines and rejected the positions 
where there were problems. It is important to note that ions with only 
one line are considered less reliable because their width $d\lambda$ 
could not be checked using this method.

\section{Ion Temperatures}
\label{sec4}

\begin{figure}
\includegraphics[width=5.7cm,angle=90]{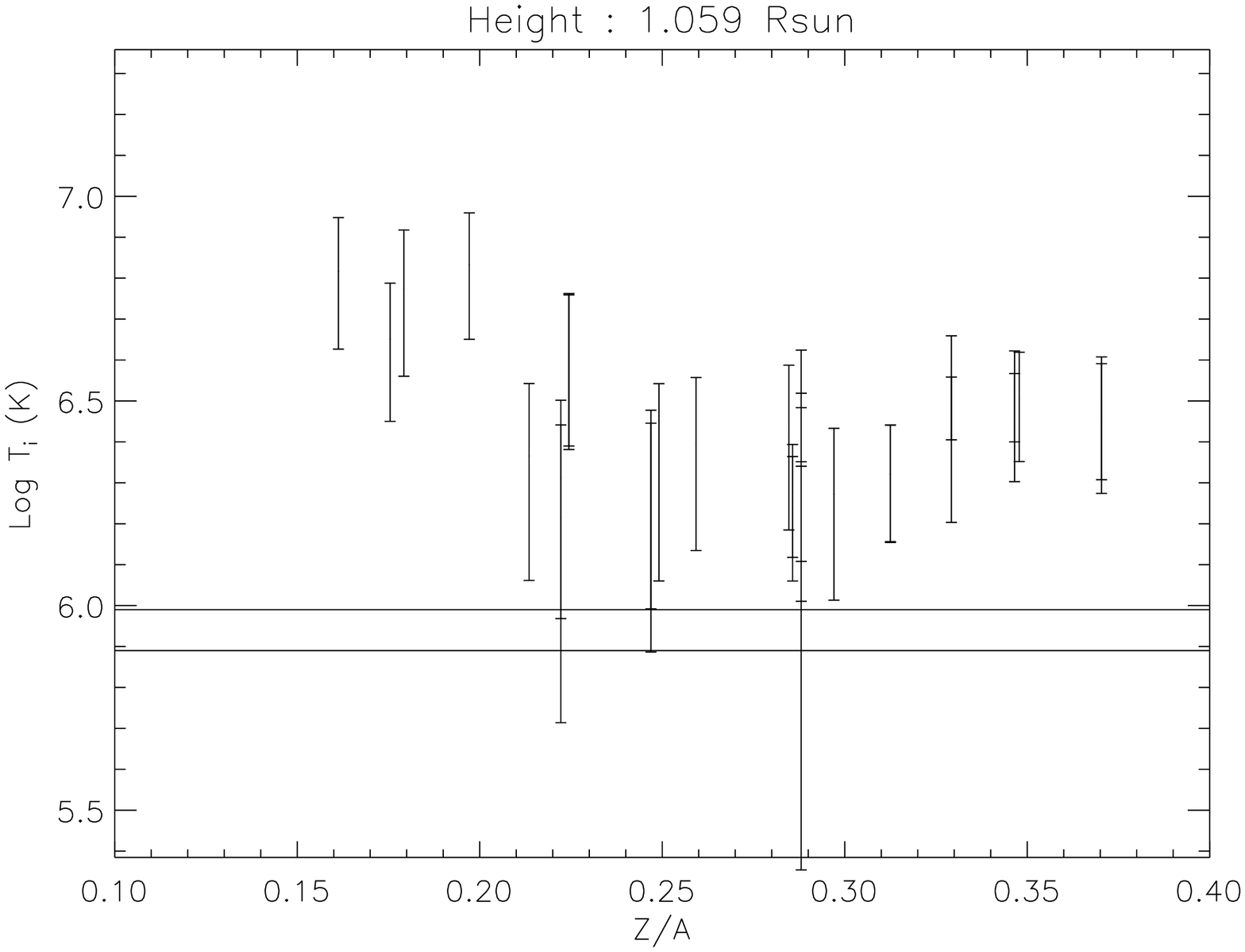}
\includegraphics[width=5.7cm,angle=90]{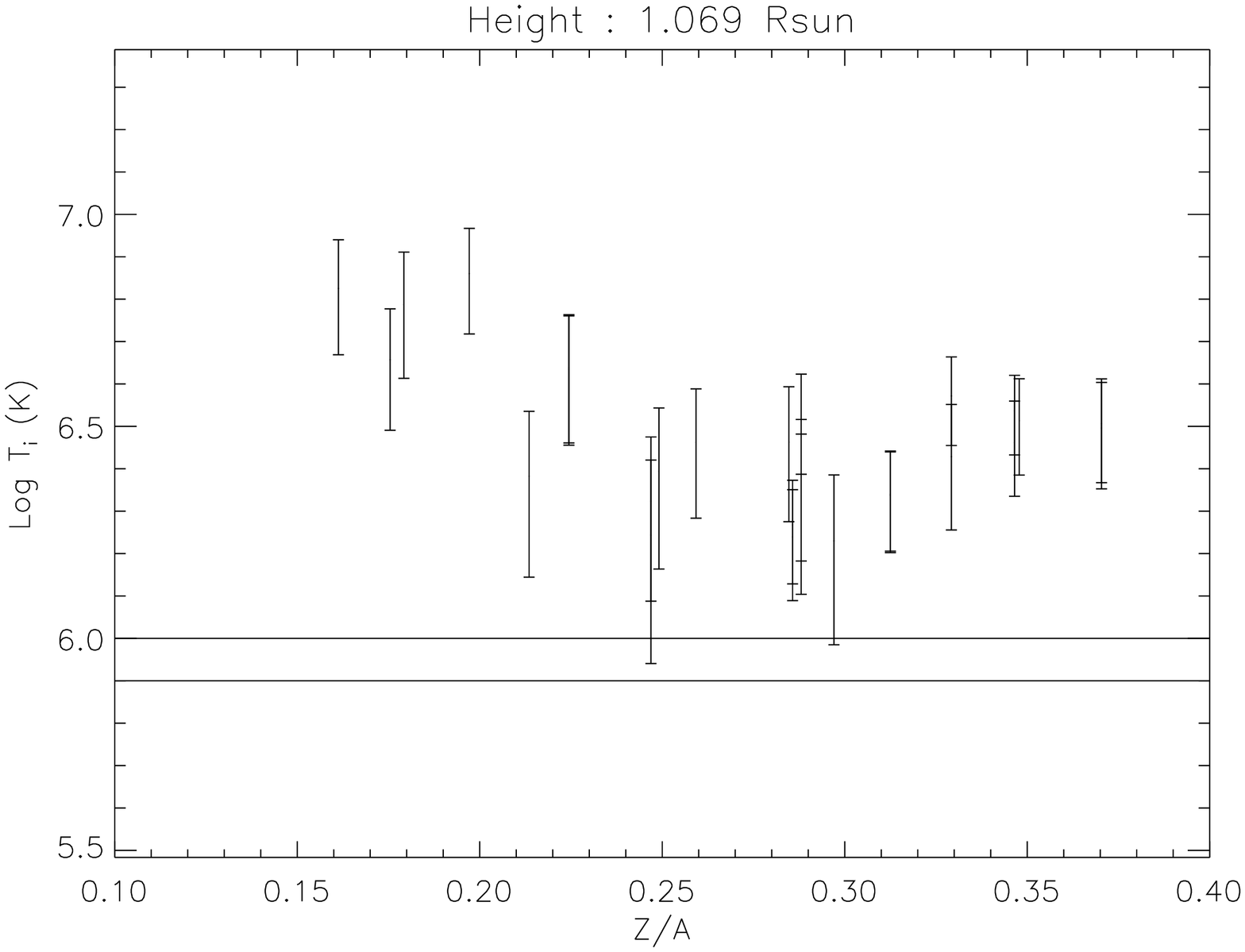}
\includegraphics[width=5.7cm,angle=90]{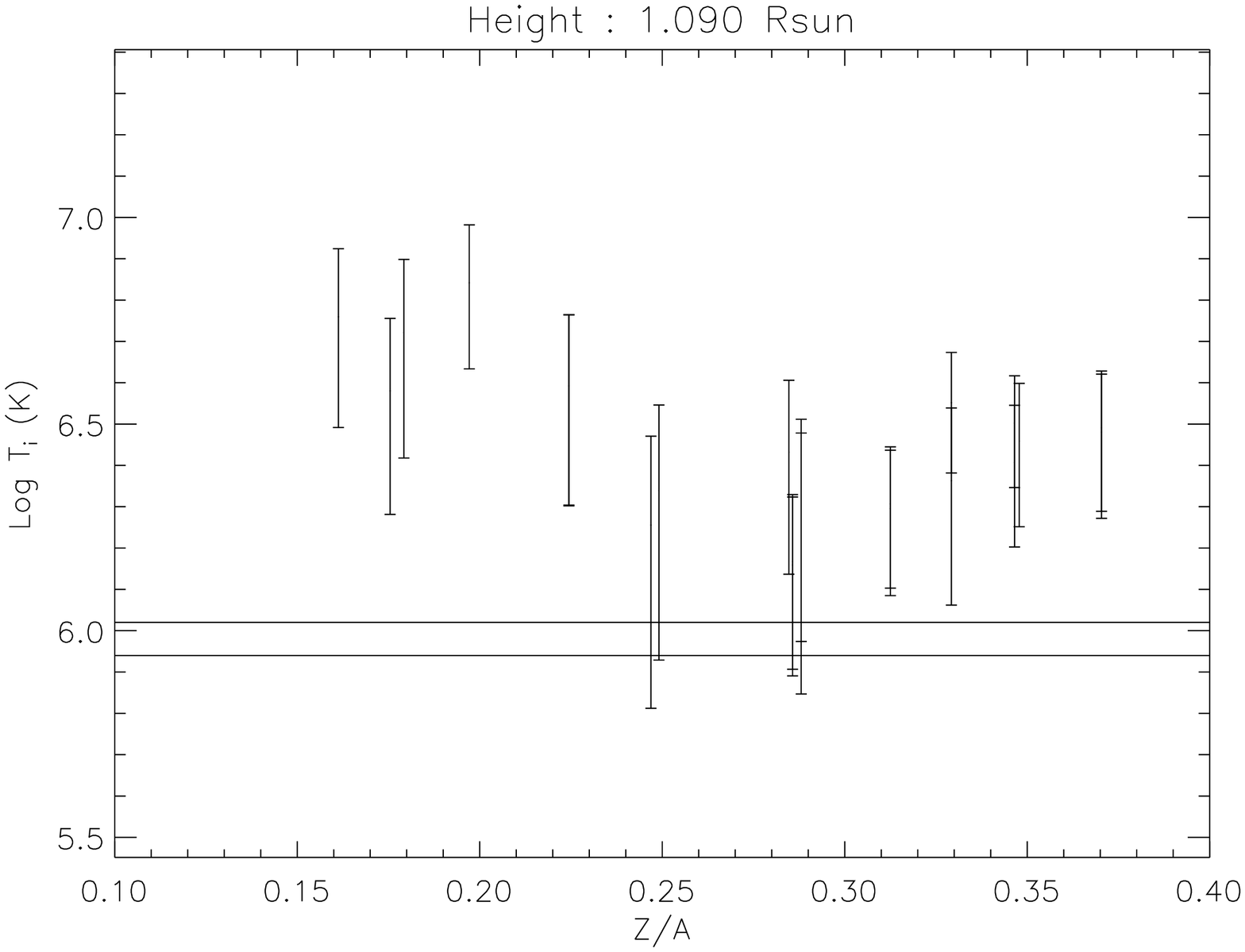}
\includegraphics[width=5.7cm,angle=90]{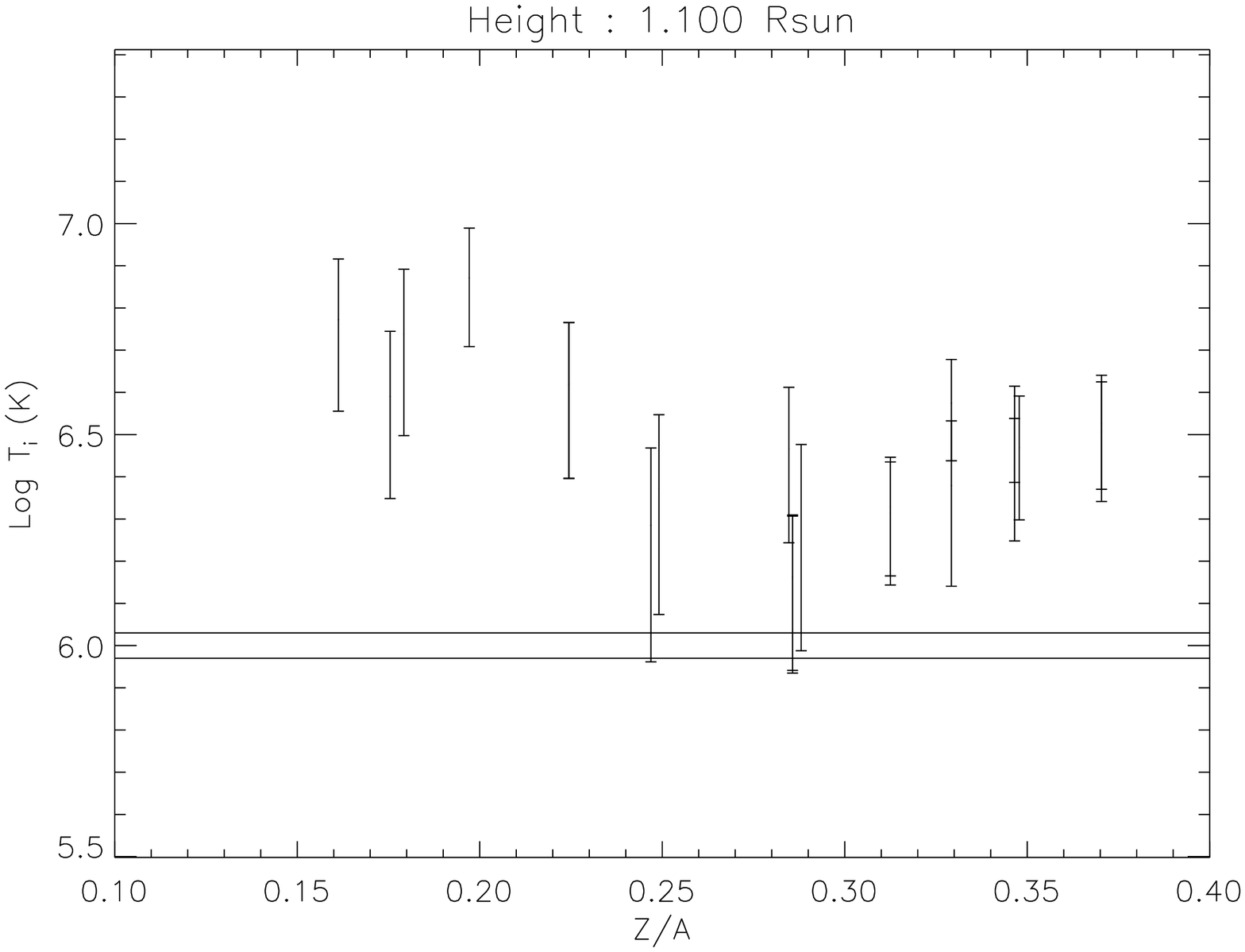}
\caption{\label{res1}
Ion temperature versus the charge-over-mass ratio $Z_{i}/A_{i}$,
for a selected subset of measured heights. The horizontal lines
represent the minimum and maximum value of the electron temperature
as measured by Landi (2008).}
\end{figure}

The ion temperatures measured from line widths are shown in Figure~\ref{res1},
which displays them as a function of the $Z_{i}/A_{i}$
ratio for a few heights above the limb.
The analysis method provides rather wide ranges of temperature for 
almost all ions, so that any height dependence is nearly washed away and the
results are qualitatively the same up to the maximum height 1.17 $R_{\odot}$. 
The ions emitting weak lines only provide a few measurements close to the 
limb, and in all cases they are compatible with a constant $T_i$ value even 
if their $T_i$ ranges seem to decrease with height. Brighter lines provide 
$T_i$ measurements up to 1.17 $R_{\odot}$, that sometimes seem to increase and 
sometimes seem to decrease with height; in all cases they too are compatible 
with constant $T_i$. As height increases, fewer and fewer lines are useful to
measure $T_i$, so that it is more difficult to sample the $Z_{i}/A_{i}$
range as accurately as with observations closer to the limb.

Landi (2008) measured the electron temperature of the same emitting plasma 
under consideration in the present work, finding that it is nearly isothermal
at each height, 
with temperature slightly increasing with height from
$\log (T_{e}/ \mbox{K} ) \simeq 5.90$ 
to $\log (T_{e}/ \mbox{K} ) \simeq 6.05$.
Figure~\ref{res1} shows the measured upper and 
lower limits of $T_e$ on top of the $T_i$ ranges: in all cases the latter 
are larger than the former and if agreement is found, it is only because 
the $T_i$ ranges are rather broad. This behavior is apparent even at the 
lowest heights and it confirms the earlier results by Tu \etal (1998): 
coronal hole plasmas are not in equilibrium.

Figure~\ref{res1} shows that there is a relationship between $T_i$
and the $Z_{i}/A_{i}$ ratio. At larger heights the number of available ions 
is limited so the relationship between $T_i$ and $Z_{i}/A_{i}$
is less apparent, or it can not even be determined,
although results are consistent with those at lower heights.
Figure~\ref{res1} shows an U-shaped $T_i$ versus $Z_{i}/A_{i}$ relationship,
where the ions providing large $T_i$ at $Z_{i}/A_{i} \leq 0.2$ are \ion[Fe x] 
to \ion[Fe xii] and \ion[Ar viii]. For $Z_{i}/A_{i} > 0.2$,
$T_i$ values are more constant, showing only a mild increase with
$Z_{i}/A_{i}$. It is 
interesting to note that all the ions with $T_i$ closer to, or in 
agreement with, $T_e$ all have $Z_{i}/A_{i}$ between 0.2 and 0.3. The 
relationship between $T_i$ and $Z_{i}/A_{i}$ can be compared with the
measurements carried out by Landi (2007) on three different quiet 
Sun regions using the same diagnostic method. Differences are striking,
as quiet Sun $T_i$ values show no dependence on the $Z_{i}/A_{i}$ ratio,
and they are in much better agreement with the electron temperature than
in the present work. However, we found that the lines we selected have 
a sufficiently large signal-to-noise to provide uncertainties in the 
limit values of $T_i$ that are small when compared to the range of 
$T_i$ given by equation~(\ref{fwhm_4}).

\begin{figure}
\includegraphics[width=6.4cm,angle=90]{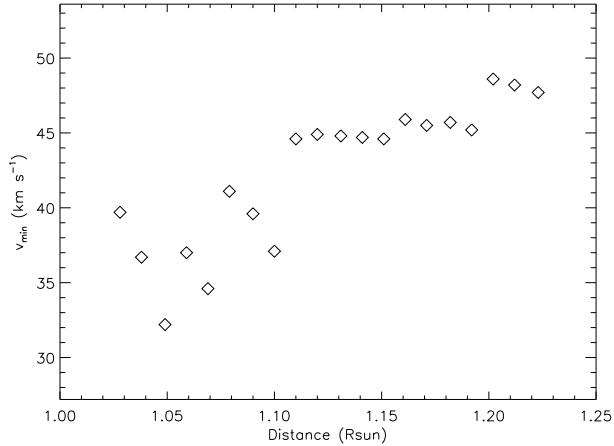}
\caption{\label{res2} Value of $v_{\rm min}$ as a function of height above
the limb.}
\end{figure}

The Tu \etal (1998) diagnostic technique does not allow us to measure 
$v_{\rm nth}$, since equation~(\ref{fwhm_3}) only allows us to set an upper
limit on it. The value of $v_{\rm min}$ increases only slightly with height 
from $\simeq$ 30--35 km s$^{-1}$ to $\simeq$ 45--50 km s$^{-1}$,
and it is shown in Figure~\ref{res2}.

\section{Thermal and Non-Thermal Components}
\label{sec5}

As discussed above, the basic assumption in the interpretation of
the off-limb line profiles is that there are only two contributors
to the (mainly Gaussian) widths:
``thermal'' microscopic motions and ``non-thermal'' bulk motions
along the line of sight (LOS) due to unresolved waves or turbulence.
Another key assumption is that the non-thermal part of the line width
has no intrinsic dependence on ion charge or mass.
This is valid when the spectrum is dominated by low-frequency MHD
waves, but other types of dispersive waves have been suggested
to play a role (e.g., Moran 2002; Ofman et al.\  2005).
However, we retain the standard assumption of no charge or mass
dependence of the non-thermal speed, because in most models of
MHD turbulence the power that resides at kinetic dispersive scales
tends to be much smaller than that at the larger non-dispersive
scales.

\begin{figure}
\epsscale{1.13}
\plotone{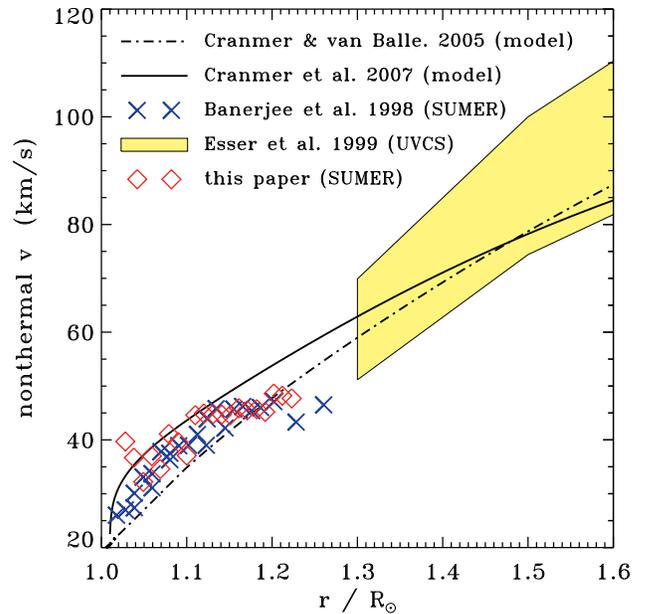}
\caption{Assembled non-thermal velocities, interpreted as perpendicular
Alfv\'{e}n wave amplitudes and plotted versus heliocentric distance in
units of the solar radius, from empirical non-WKB wave models
({\em dot-dashed line;} Cranmer \& van Ballegooijen 2005) and
turbulence-driven coronal heating models ({\em solid line;}
Cranmer et al.\  2007).
Also, empirical measurements from Banerjee et al.\  (1998)
({\em blue crosses}), Esser et al.\  (1999)
({\em yellow region}), and from this paper ({\em red diamonds}).}
\label{steve1}
\end{figure}

Figure~\ref{steve1} compares a range of observational and theoretical values
for the non-thermal velocity, including the upper limits derived above.
This quantity is interpreted as the transverse root mean square
Alfv\'{e}n wave amplitude $\langle \delta v_{\perp}^{2} \rangle^{1/2}$.
The wave amplitudes from theoretical models were divided by
$\sqrt{2}$ in order to sample only the motions along one of the
two perpendicular directions (i.e., only along the LOS).
Cranmer \& van Ballegooijen (2005) discussed the observational
data of Banerjee et al.\  (1998) and Esser et al.\  (1999),
as well as other empirical constraints on Alfv\'{e}n wave properties.
The discrepancy between the two theoretical model curves is due to
the fact that the Cranmer \& van Ballegooijen (2005) model utilized
a fixed empirical set of background plasma parameters (e.g., wind speed,
Alfv\'{e}n speed, density) and Cranmer et al.\  (2007) computed these
properties self-consistently with the waves.

Because in coronal holes (close to the limb) the magnetic field
appears to be mainly pointed in the radial direction, the LOS
component of the thermal temperature (for ion $i$) is assumed
to be identical to the perpendicular temperature $T_{\perp i}$.
This assumption was validated by test calculations that took account
of the full LOS integration and used anisotropic bi-Maxwellian
distributions.
These calculations showed that the dominant contribution to a
collisional line profile (i.e., sensitive to density squared) comes
from a very narrow range of heights, all close enough to the plane of
the sky that the sampled component of the distribution always remained
close to $T_{\perp i}$.

The model calculations below mainly utilize the ion temperature
data at a single reference radius.
This radius was chosen as 1.069 $R_{\odot}$, which seemed to be
the best balance between having a large number of available ions
(which drops off as height increases) and the existence of
significant ``preferential heating'' away from thermal
equilibrium (which is lessened as height decreases).
The measured maximum non-thermal speed at this height is
34.6 km s$^{-1}$, which falls between the modeled values of about
30 km s$^{-1}$ (Cranmer \& van Ballegooijen 2005) and 40 km s$^{-1}$
(Cranmer et al.\  2007).
However, the actual non-thermal speed must be less than the upper
limit derived above.
Somewhat arbitrarily, we choose 25 km s$^{-1}$ as a fiducial
non-thermal speed for this height, which is used to subtract from
the total line width to obtain the individual ion temperatures
$T_{\perp i}$.\footnote{%
To some extent, when comparing the ion temperatures
{\em to one another} at the same height, the assumed value for the
non-thermal wave amplitude is not so important.
As long as the non-thermal speed is independent of ion charge and mass,
the {\em relative} differences between the various ion temperatures
are unchanged.}

\section{Physical Processes in the Model}
\label{sec6}

The basic idea of the semi-empirical models to be discussed below
is that we assume a turbulent replenishment of ion cyclotron
wave power in the corona with a known radial dependence.
The absolute normalization for this wave power, though, is determined
for a given ion of charge $Z_i$ and mass $A_i$
iteratively by finding the value that produces the observed level of
perpendicular heating at the reference radius of 1.069 $R_{\odot}$.
By determining this wave power for the ions, each independently of the
others, we construct a model of the overall frequency dependence of the
ion cyclotron power spectrum.

The solution for the radial dependence of $T_{\perp i}$ uses an
internal energy conservation equation similar to that used by
Cranmer et al.\  (1999a) with some minor changes to be discussed
below.
The properties of the background proton-electron plasma
are held fixed at values determined by comparison with empirical
measurements (mainly those of Cranmer \& van Ballegooijen 2005).
The heavy ions are thus treated as ``test particles'' that do not
influence the primary plasma or one another.
A dominant effect that needs to be included is the temperature
equilibration due to Coulomb collisions between protons and
the heavy ion species of interest.
The high densities in the low corona give rise to rapid collisions
that can mask the preferential ion heating from ion cyclotron waves
(by driving $T_{\perp i}$ back down to the proton temperature $T_p$).

The time-steady energy conservation equation that we solve is
given by
\begin{equation}
  \frac{d T_{\perp i}}{dr} +
  \frac{T_{\perp i}}{\cal A} \frac{d{\cal A}}{dr} \, = \,
  \frac{Q_{\perp i} + C_{\perp ip} (T_{\perp p} - T_{\perp i})}
  {n_{i} u_{i} k_{\rm B}}
  \label{eq:dTperp}
\end{equation}
where the two terms on the right-hand side represent resonant heating
($Q_{\perp i}$) and Coulomb collisions with protons ($C_{\perp ip}$).
The second term on the left-hand side accounts for adiabatic expansion,
and depends on the radial variation in the cross-sectional area
${\cal A}$ of the polar flux tube.
One notable departure from the models of Cranmer et al.\  (1999a)
is that we do not consider the corresponding equation for the parallel
ion temperature $T_{\parallel i}$.
Thus, for simplicity, we assume that there is no significant dependence
of $T_{\perp i}$ on $T_{\parallel i}$.
Even for strong anisotropy, the isotropization (coupling) term
$J_{\perp ip}$ from Cranmer et al.\  (1999a) is a factor of
$\sim 0.2/A_{i}$ weaker than the standard proton-ion collision term,
and thus is negligible.

The proton-ion collision coefficient in equation (\ref{eq:dTperp})
is given by $C_{\perp ip} = 2 k_{\rm B} n_{i} \nu_{ip}$, where
$n_i$ is the ion number density. 
The Coulomb collision rate itself is
\begin{equation}
  \nu_{ip} \, = \,
  \frac{16}{3} \pi^{1/2} \ln \Lambda \,
  \frac{q_{i}^{2} q_{p}^{2} n_p}{m_{i} m_{p} a_{\parallel}
  a_{\perp}^2}
\end{equation}
(e.g., Spitzer 1962), and 
\begin{equation}
  a_{\parallel} \, \equiv \, \left[
  \left( 2 k_{\rm B} T_{\parallel i} / m_i \right) +
  \left( 2 k_{\rm B} T_{\parallel p} / m_p \right) \right]^{1/2}
\end{equation}
\begin{equation}
  a_{\perp} \, \equiv \, \left[
  \left( 2 k_{\rm B} T_{\perp i} / m_i \right) +
  \left( 2 k_{\rm B} T_{\perp p} / m_p \right) \right]^{1/2}
\end{equation}
(see also Barakat \& Schunk 1981, 1982; Isenberg 1984). 
As in Cranmer et al.\  (1999a), we take the Coulomb logarithm to
be a constant of $\ln \Lambda = 21$.
We assume the protons are isotropic, and we also set
$T_{\parallel i} = T_p$ in the definition of $a_{\parallel}$
(which was verified to hold at the SUMER heights in trial runs
using the original equations from Cranmer et al.\  1999a).
We do not consider collisional interactions between ion species $i$
and any other heavy ion species, which implies that ion $i$ does not
ionize or recombine to any other charge states over the range of heights 
(see, e.g., Lie-Svendsen \& Esser 2005).

The resonant cyclotron heating rate is given in the so-called
``optically thin'' (narrow resonance zone) limit of Cranmer (2000) as
\begin{equation} 
  \frac{Q_{\perp i}}{m_{i} n_i} \, = \,
  \frac{4 \pi^{2} \, {\cal P} (\Omega_{i}, r)}{B_{0}^2} \,\,
  \Omega_{i}^{2} V_{\rm A}^{2}
  \left( 1 - \frac{Z_i}{A_i} \right)
  \label{eq:Qzone} 
\end{equation}
where $B_0$ is the radial background magnetic field strength,
$\Omega_i$ is the ion gyrofrequency, and $V_{\rm A}$ is the
Alfv\'{e}n speed.
The resonant wave power spectrum ${\cal P} (\omega)$ differs from
the power spectrum used by Cranmer et al.\  (1999a) by a factor of
$4\pi$ because the latter was defined as the spectrum of the
total magnetic fluctuation variance $\langle \delta B_{\perp}^{2} \rangle$.
${\cal P} (\omega)$, however, is defined as the spectrum of the total
fluctuation energy density (see eqs.~[\ref{eq:P3Dint}] and
[\ref{eq:Pconv}] below).

Cranmer et al.\  (1999a) assumed that the power spectrum scaled with
the overall radial dependence of $\langle \delta B_{\perp}^{2} \rangle$,
as computed via MHD wave action conservation.
However, the relatively small amount of energy in the high-frequency
tail of the spectrum may not necessarily scale with the total
integrated power.
Thus, here we modify this radial dependence using the MHD turbulence
theory of Cranmer \& van Ballegooijen (2003) to take account of an
assumed anisotropic cascade that maintains the high-frequency tail.
We assume that ${\cal P} (\Omega_{i},r)$ is a separable function
of radius and frequency (for which $\omega \approx \Omega_{i}$).
The radial dependence is given by
\begin{equation}
  {\cal P} (\Omega_{i}, r) \, \propto \,
  \frac{V_{\rm A}^{3/2}}{\rho^{1/2} (u_{p} + V_{\rm A})^3}
  \,\, ,
  \label{eq:resrad}
\end{equation}
where $\rho$ is the mass density.
This scaling relation is derived in Section~\ref{sec7} below, and the
normalization constant to be used is the main adjustable parameter
that we modify to produce agreement with the measured values of
$T_{\perp i}$ at 1.069 $R_{\odot}$.

The background plasma properties are given by the empirical model
of Cranmer \& van Ballegooijen (2005).
These properties include the polar flux-tube area ${\cal A}(r)$, the
outflow speeds (for which we assume that the ions are flowing at the
same speed as the protons; $u_{i} = u_{p}$), and the number densities.
The primary plasma number density ($n_{p} \approx n_{e}$) is assumed to
have the same radial dependence as the ion number density $n_i$.
Note that the absolute value of the ion number density is never needed,
only its radial dependence.

The Cranmer \& van Ballegooijen (2005) model was essentially a
``cold plasma'' treatment, so our description of the background proton
temperature $T_p$ is given by scaling the self-consistent polar coronal
hole temperature of Cranmer et al.\  (2007) up and down by arbitrary
constant factors.
The location of the transition region was also scaled down to a standard
height of 0.003 $R_{\odot}$ above the photosphere.
We assume isotropic protons with
$T_{\parallel p} = T_{\perp p} \equiv T_{p}$.
The actual proton temperature in the corona is not well constrained
by observations, since line ratio techniques are sensitive to the
electron temperature, and the widths of H$^0$ lines are more strongly
``contaminated'' by non-thermal motions than the lines of heavy ions.

\begin{figure}
\epsscale{1.10}
\plotone{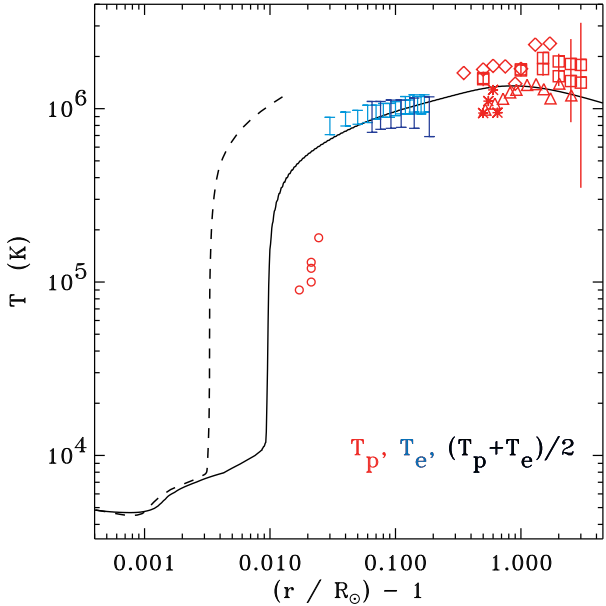}
\caption{Assembled temperatures in polar coronal holes:
Mean plasma temperatures from a semi-empirical VAL/FAL-type
model ({\em dashed line;} Avrett \& Loeser 2008) and from a
turbulence-driven coronal heating model ({\em solid line;}
Cranmer et al.\  2007).
$T_e$ from off-limb SUMER measurements made by Wilhelm (2006)
({\em dark blue bars}) and Landi (2008) ({\em light blue bars}).
$T_p$ from off-limb SUMER measurements by
Marsch et al.\  (2000) ({\em circles}), and UVCS measurements
assembled by Cranmer (2004), with data from
Cranmer et al.\  (1999b) ({\em squares}),
Esser et al.\  (1999) ({\em diamonds}),
Zangrilli et al.\  (1999) ({\em asterisks}), and
Antonucci et al.\  (2000) ({\em triangles}).
\hspace*{0.3in} \hspace*{0.3in} \hspace*{0.3in}
\hspace*{0.3in} \hspace*{0.3in} 
\hspace*{0.3in} \hspace*{0.3in}}
\label{steve2}
\end{figure}

In order to determine what to use for the proton temperature in the
models, Figure~\ref{steve2} shows a range of measurements and model
predictions for $T_{p}$, as well as the electron temperature $T_e$ (which
is expected to be collisionally coupled to the protons at high enough
densities) and the mean one-fluid temperature, which should be
approximately proportional to $(T_{p} + T_{e})/2$.
Off-limb measurements of $T_e$ in polar coronal holes are shown from
SUMER data described by Wilhelm (2006) and Landi (2008).
Proton temperatures at large heights have been measured by
UVCS/{\em{SOHO}}, and the plotted points are those assembled by
Cranmer (2004), with a model prediction for the non-thermal wave
amplitudes subtracted from the empirical kinetic temperatures.
The transition-region-like values of $T_p$ at low heights were
measured by Marsch et al.\  (2000); these are quite low compared to
most time-steady coronal models that place the sharp transition region
at a lower height around $\approx$ 0.003 $R_{\odot}$.
These observations may have been influenced by the prevalence of
narrow and cool spicules above the limb and may not be representative
of the coronal footpoints of solar wind streams.

\begin{figure}
\epsscale{1.13}
\plotone{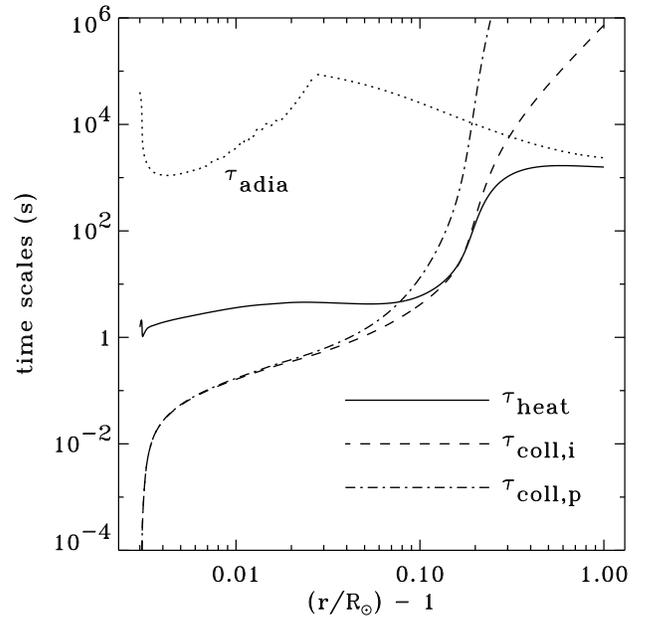}
\caption{Radial dependence of timescales corresponding to individual
terms in the energy conservation equation (\ref{eq:dTperp}); see
text for definitions.}
\label{stevePOSTREF}
\end{figure}

Before proceeding to discuss the solution of equation (\ref{eq:dTperp}),
it is important to make clear the relative strengths of the
individual heating and cooling terms in this conservation equation.
As an illustration, Figure~\ref{stevePOSTREF} shows the radial
dependence of {\em timescales} computed from each term in
equation (\ref{eq:dTperp}) for the standard choice of parameters
(and the O$^{5+}$ ion) discussed in {\S}~8.
Given that the terms in this equation have the units of a spatial
temperature gradient, a timescale can be constructed by multiplying
by the quantity $u_{i}/T_{\perp i}$ and taking the inverse.
For example, the timescale corresponding to ion cyclotron heating
is given by
\begin{equation}
  \tau_{\rm heat} \, = \,
  \frac{n_{i} k_{\rm B} T_{\perp i}}{Q_{\perp i}}  \,\, .
\end{equation}
The timescale for adiabatic cooling $\tau_{\rm adia}$ is given by
using the second term on the left-hand side of
equation (\ref{eq:dTperp}).
Two timescales for Coulomb collisions can be constructed by using
either the ion temperature (i.e., the final term in eq.~[\ref{eq:dTperp}])
or the proton temperature (the pentultimate term).
These generate the collisional timescales $\tau_{{\rm coll,}i}$ and
$\tau_{{\rm coll,}p}$, respectively.
As can be seen from Figure~\ref{stevePOSTREF}, at low heights
the most rapid processes are the Coulomb collisions.
Above a height of approximately 0.1 $R_{\odot}$, the ion cyclotron
heating timescale begins to be as rapid as collisions, and the
ion temperature $T_{\perp i}$ begins to increase above the proton
temperature.
At still larger heights in the extended corona ($r \approx 2 R_{\odot}$)
adiabatic cooling begins to be as important as ion cyclotron heating,
and Coulomb collisions are unimportant.

The numerical integration technique used to compute $T_{\perp i}(r)$
from equation (\ref{eq:dTperp}) was slightly different from that used
by Cranmer et al.\  (1999a).
Because of the strong collisions near the coronal base, the earlier
use of simple Euler steps became highly unstable for realistic grid
spacings.
Instead, we used a piecewise analytic solution of the energy equation
that takes account of the collisional coupling explicitly.
Equation (\ref{eq:dTperp}) can be expressed schematically as
\begin{equation}
  \frac{d T_{\perp i}}{dr} + p T_{\perp i} \, = \, q
\end{equation}
where $p$ and $q$ are considered to be slowly varying functions of
radius.
The Coulomb collision term has been split up between $p$ and $q$.
The analytic solution of this equation, which we use to step the
solution for $T_{\perp i}$ up from height $r_j$ to height $r_{j+1}$,
is given by the integrating factor method as
\begin{equation}
  T_{\perp i, j+1} \, = \, \frac{q}{p} + e^{-p(r_{j+1} - r_{j})}
  \left( T_{\perp i, j} - \frac{q}{p} \right)
  \,\, .
  \label{eq:ifac}
\end{equation}
In the limit of weak collisions ($p \ll 1$), this reduces to the 
standard Euler step,
\begin{equation}
  T_{\perp i, j+1} \, \approx \, 
  T_{\perp i, j} + q (r_{j+1} - r_{j})  \,\, .
\end{equation}
In the limit of strong collisions (i.e., ignoring the adiabatic
expansion and cyclotron heating terms), the ratio $q/p$ reduces to
the proton temperature $T_p$, and equation (\ref{eq:ifac}) is seen
to stably drive the solution towards thermal equilibrium.

\section{Expectations from Anisotropic Turbulence Theory}
\label{sec7}

As an independent prediction of the resonant wave power, we aim to
describe the anisotropic evolution of wave energy in 2D wavenumber
space ($k_{\parallel}, k_{\perp}$) by means of the advection-diffusion
theory described by Cranmer \& van Ballegooijen (2003) and in
Appendix C of Cranmer \& van Ballegooijen (2005).
The theoretical development given in this section is somewhat
independent of the ``semi-empirical'' derivation of wave power
that comes from solving the equations of the previous section.

The goal is to develop an analytic solution for the turbulent 
power spectrum (for frequencies near the ion cyclotron resonances)
in the limit that the cascade is allowed to proceed to its final
``driven'' steady state in a small homogeneous volume of plasma.
The three-dimensional total power spectrum is defined as
\begin{equation}
  \int d^{3} {\bf k} \, P_{\rm 3D} ({\bf k})
  \, = \, U_{\rm tot} \, \approx \,
  \frac{\langle \delta B_{\perp}^{2} \rangle}{4\pi}
  \label{eq:P3Dint}
\end{equation}
where $U_{\rm tot}$ is the total energy density of the Alfv\'{e}n
waves, and we write the volume element $d^{3} {\bf k}$ in cylindrical
coordinates as $2\pi k_{\perp} \, dk_{\perp} dk_{\parallel}$
to assume symmetry in the two directions perpendicular to
the background field.

For fully developed anisotropic MHD turbulence, we assume the
power spectrum to be a separable function of two variables:
$k_{\perp}$ and a nonlinearity parameter $y$ defined as the
ratio of the local wind-frame frequency $V_{A} k_{\parallel}$ to
an assumed nonlinear eddy turnover rate
$\langle \delta V \rangle k_{\perp}$ (see, e.g.,
Goldreich \& Sridhar 1995).
We use the notation from Section~2.3 of
Cranmer \& van Ballegooijen (2003) and define
\begin{equation}
  P_{\rm 3D} (k_{\parallel}, k_{\perp}) \, = \,
  \frac{\rho V_{A} W_{\perp}^{1/2}}{k_{\perp}^3} \, g(y)
  \label{eq:P3D}
\end{equation}
where $W_{\perp} (k_{\perp})$ is a reduced power spectrum that
describes the dominant perpendicular cascade.
For the MHD inertial range, the reduced spectrum is given as
\begin{equation}
  W_{\perp} (k_{\perp})  =  \left\{
  \begin{array}{ll}
    U_{\rm tot} ( k_{\perp} / k_{\rm out} )^{-2/3}
      (3\pi\rho)^{-1} \, ,
      & k_{\rm out} < k_{\perp} < k_{\rm in} \\
    0  \, , & \mbox{otherwise} \\
  \end{array} \right.
  \label{eq:wperpout} .
\end{equation}
The above definition assumes the existence of a finite
outer-scale perpendicular wavenumber $k_{\rm out}$, which we
assumed to be inversely proportional to the correlation length of the
turbulence, and an inner-scale wavenumber $k_{\rm in} \gg k_{\rm out}$,
which we assume to be equivalent to the inverse proton gyrofrequency.
The factor of $3 \pi \rho$ above is needed to normalize the
full power spectrum as defined in eq.~(\ref{eq:P3Dint}).

The $k_{\parallel}$ dependence of the power spectrum is
contained in the dimensionless $g(y)$ function given in
eq.~(\ref{eq:P3D}).
We define
\begin{equation}
  y \, = \, \frac{k_{\parallel} V_A}{k_{\perp} W_{\perp}^{1/2}}
  \,\, .
\end{equation}
The condition $y=1$ is defined as ``critical balance'' by
Goldreich \& Sridhar (1995), and their analysis only
constrains the general shape of $g(y)$, not its exact value.
The critical balance condition captures the highly nonlinear state
of turbulence, for which a coherent wave survives for no more
than about one or two periods before nonlinear processes transfer
its energy to the smaller scales.
Cranmer \& van Ballegooijen (2003) solved a simple wavenumber
diffusion equation to obtain an analytic relation for $g(y)$.
For the MHD inertial range, this relation can be shown to be
equivalent to
\begin{equation}
  g(y) \, = \, \frac{2 \Gamma (n)}{3 \Gamma(n - 0.5) \sqrt{\pi}}
  \left( 1 + \frac{4 y^{2}}{9} \right)^{-n}
  \label{eq:gkappa}
\end{equation}
which is normalized to unity when integrated over all $y$, and with
\begin{equation}
  n \, = \, 1 + \frac{3 \beta}{4 \gamma}
  \,\, .
\end{equation}
The dimensionless constants $\beta$ and $\gamma$ describe the relative
strengths of advection and diffusion, respectively, in the $k_{\perp}$
direction.
They only occur as the ratio $\beta / \gamma$, which we take to
be a free parameter.
Cranmer \& van Ballegooijen (2003) discussed the most realistic
values for this ratio, which is relatively unconstrained by
existing turbulence simulations.
The ``random walk'' turbulence model of van Ballegooijen (1986),
though, suggested that $\beta / \gamma \approx 1$.
However, Cranmer \& van Ballegooijen (2003) found that one would
need this ratio to be smaller than about 0.25 in order to produce
enough parallel cascade in the corona to heat protons and heavy ions
via cyclotron resonance.

With the above definitions it becomes possible to integrate over
$k_{\perp}$ and obtain a reduced 1D power spectrum
\begin{equation}
  P_{\rm 1D} (k_{\parallel}) \, = \,
  2\pi \int_{k_{\rm out}}^{k_{\rm in}} dk_{\perp} \, k_{\perp}
  \, P_{\rm 3D} (k_{\parallel},k_{\perp})
\end{equation}
which is related to the frequency spectrum ${\cal P}(\omega)$
via
\begin{equation}
  P_{\rm 1D} (k_{\parallel}) dk_{\parallel} \, = \, 
  {\cal P} (\omega) d\omega  \,\, .
  \label{eq:Pconv} 
\end{equation}
If we assume, for simplicity, that the ion cyclotron waves of interest
are far enough from the proton resonance for the standard MHD
dispersion relation ($\omega = k_{\parallel} V_{\rm A}$) to hold,
the above parameterizations for $P_{\rm 3D} (k_{\parallel},k_{\perp})$
can be applied to show that
\begin{equation}
  {\cal P} (\omega) \, = \, \sqrt{\frac{4\pi\rho U_{\rm tot}}{3}}
  k_{\rm out}^{1/3} \int_{k_{\rm out}}^{k_{\rm in}} dk_{\perp}
  \frac{g(y)}{k_{\perp}^{7/3}}
  \,\, .
  \label{eq:Pomega}
\end{equation}
This equation, combined with the cyclotron resonant frequency condition
\begin{equation}
  \omega - u_{i} k_{\parallel} - \Omega_{i} \, = \, 0  \,\, ,
\end{equation} 
is used to produce the ``theoretical'' power spectra (as a function of
the $\beta / \gamma$ parameter) that are compared with the empirically
derived power spectra at 1.069 $R_{\odot}$. 
For ion cyclotron waves at any given height in the low corona
(where $u_{i} \ll V_{\rm A}$), the wave frequency $\omega$ is
directly proportional to the ion charge to mass ratio
$Z_{i}/A_{i}$.
The main plasma quantities, including the Alfv\'{e}n wave scale
lengths $k_{\rm out}$ and $k_{\rm in}$, the total energy density
$U_{\rm tot}$, and the ion Larmor frequencies $\Omega_i$, are
taken from the radially dependent coronal hole model of
Cranmer \& van Ballegooijen (2005).

The above equations specify how the resonant frequency spectrum
${\cal P}(\omega)$ can be computed, but it is also helpful to
derive a concise relation for the large-scale radial dependence of the
wave power in the ion cyclotron frequency range.
The gyroresonant wave power exists in the limit of large values of
the dimensionless critical-balance parameter $y$.
For $y \gg 1$, the function $g(y) \propto y^{-2n}$, and this implies
that the high-frequency tail of the spectrum has a similar power-law
dependence ${\cal P} \propto \omega^{-2n}$.
Cranmer et al.\  (1999) treated this exponent as a free parameter
$\eta$.
Utilizing the definitions given above, this proportionality can be
expressed in terms of more basic plasma parameters as
\begin{equation}
  g(y) \, \propto \, k_{\perp}^{4n/3} \left(
  \frac{k_{\rm out}^{2/3} U_{\rm tot}}{3\pi\rho \Omega_{i}^2}
  \right)^{n}  \,\, .
\end{equation}
Note that the actual {\em frequency} dependence is ``hidden'' in
in the ion gyrofrequency $\Omega_i$, which is simply $Z_{i}/A_{i}$
times the proton gyrofrequency $\Omega_p$.
Performing the integration in equation (\ref{eq:Pomega}) and removing all
constants thus provides the radial dependence of the power spectrum:
\begin{equation}
  {\cal P} \, \propto \,
  \frac{\rho^{0.5 - n} U_{\rm tot}^{n + 0.5} k_{\rm out}^{(2n + 1)/3}}
  {\Omega_{p}^{2(n+2)/3} w_{p}^{4(n-1)/3}}
\end{equation}
where the proton thermal speed $w_p$ is part of the definition
of the mean proton gyroradius used in $k_{\rm in}$.
Also, the constant ratio of $Z_{i}/A_{i}$ has been removed in order
to isolate the radial dependence for any given ion resonance.

We will see below that the limiting case of $\beta / \gamma \ll 1$
(i.e., $n \approx 1$, or $\eta \approx 2$ in the notation of
Cranmer et al.\  1999) may be appropriate for comparison with the
observations, so let us apply this limit to obtain
\begin{equation}
  {\cal P} \, \propto \,
  \rho^{-1/2} U_{\rm tot}^{3/2} k_{\rm out} \Omega_{p}^{-2}
  \,\, .
\end{equation}
The outer scale wavenumber $k_{\rm out}$ is usually assumed to
be the reciprocal of the perpendicular correlation length, and the
latter is often taken to be proportional to the flux-tube cross
section, so we assume $k_{\rm out} \propto A^{-1/2} \propto
B_{0}^{1/2}$ (Hollweg 1986).
We also know that $\Omega_{p} \propto B_{0}$.
Finally, the total Alfv\'{e}n wave energy density can be assumed to
scale with the WKB conservation of wave action, which is given by
\begin{equation}
 U_{\rm tot} \, \propto \, \frac{V_{\rm A}}{{\cal A} (u_{p} + V_{\rm A})^2}
 \,\, .
\end{equation}
Applying these scalings then leads to the radial dependence for the
resonant power given in equation (\ref{eq:resrad}).

\section{Model Results}
\label{sec8}

The semi-empirical equation of ion energy conservation
(eq.~[\ref{eq:dTperp}]) was integrated numerically from the coronal
base (1.003 $R_{\odot}$) to the reference radius of 1.069 $R_{\odot}$
for each of 25 ions having measured SUMER line widths at this height.
These ions have $Z_{i}/A_{i}$ values ranging between 0.161 (Fe$^{9+}$)
and 0.370 (Mg$^{9+}$), and ion temperatures between 1.69 MK (Ne$^{6+}$)
and 7.22 MK (Fe$^{11+}$).
Also, several grids of models were created for a range of proton
temperatures, all with the same relative radial dependence but with
values at 1.069 $R_{\odot}$ ranging between 0.5 and 3 MK.

\begin{figure}
\epsscale{1.13}
\plotone{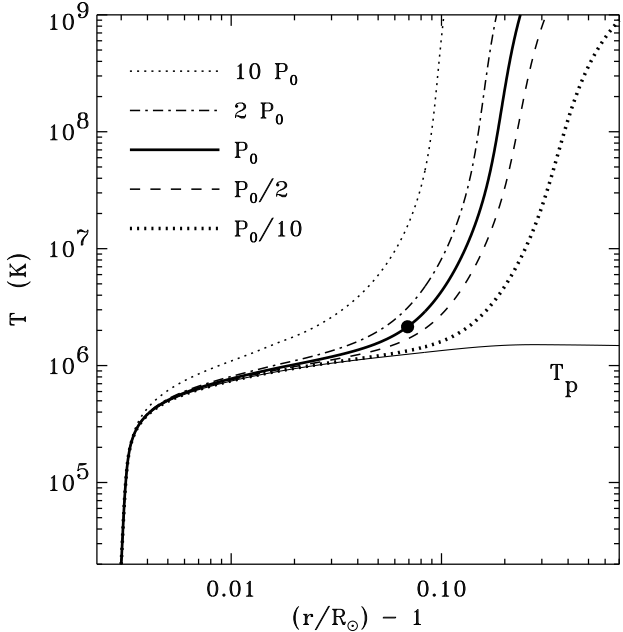}
\caption{Radial dependence of modeled proton temperature
({\em thin solid line}) and perpendicular O$^{5+}$ temperatures,
assuming a range of resonant wave power normalization values
given relative to $P_0$, the iterated value required to heat the ions
to the observed level at 1.069 $R_{\odot}$ ({\em thick solid line}).
Resulting values for $T_{\perp i}(r)$ for power levels of
$10 P_0$ ({\em thin dotted line}), $2 P_0$ ({\em dot-dashed line}),
$0.5 P_0$ ({\em dashed line}), and $0.1 P_0$ ({\em thick dotted line})
are also shown.}
\label{steve3}
\end{figure}

Figure~\ref{steve3} illustrates the result of this numerical integration for
a representative ion (O$^{5+}$) and a specific choice for the proton
temperature.
The optimal (iterated) wave power necessary to heat these ions to
the observed temperature at the reference height is denoted $P_0$.
Figure~\ref{steve3} also shows the resulting $T_{\perp i}(r)$ curves that are
obtained when $P_0$ is varied up and down by factors of 2 to 10.
The ion temperature can become several orders of magnitude larger
than the proton temperature at larger heights.
This has been confirmed by the UVCS \ion[O vi] measurements between
1.5 and 3.5 $R_{\odot}$ (e.g., Kohl et al.\  2006;
Cranmer et al.\  2008).
Note, though, that at such large heights, where Coulomb collisions
become very weak, a more proper treatment of the resonant wave-particle
interactions is warranted---either via the coupled
$T_{\perp i}$ and $T_{\parallel i}$ equations (Cranmer et al.\  1999a)
or even a fully kinetic treatment of the ion velocity distributions
(e.g., Isenberg \& Vasquez 2007).

\begin{figure}
\epsscale{1.13}
\plotone{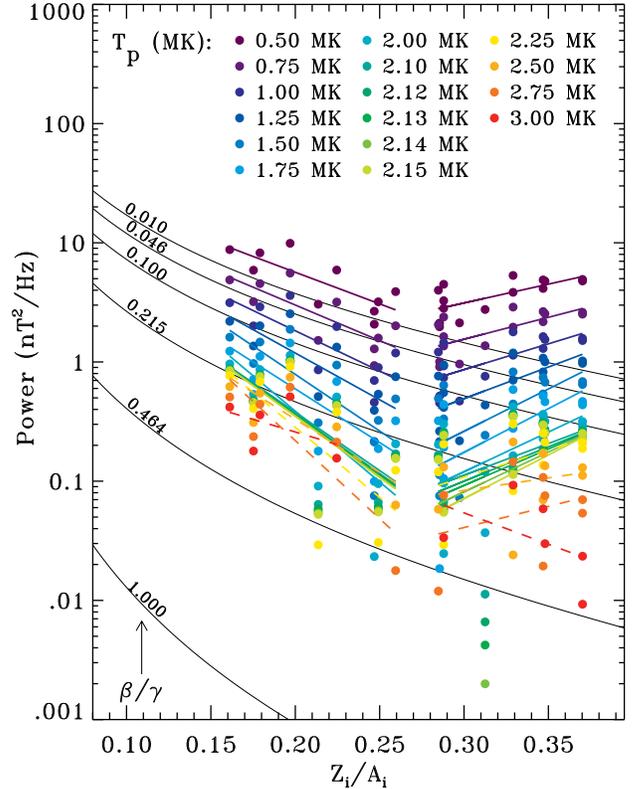}
\caption{Ion cyclotron wave power at 1.069 $R_{\odot}$ plotted
versus ion charge-to-mass ratio $Z_{i}/A_{i}$ (the latter a proxy
for frequency).
Empirically determined wave power levels for individual ions
({\em filled circles;} see color key for $T_p$ normalization)
are compared with theoretical predictions for the wave power, based
on adopted radially dependent models of the total wave power,
Alfv\'{e}n speed, and other plasma properties ({\em black
solid curves}).}
\label{steve4}
\end{figure}

Figure~\ref{steve4} shows the result of repeating the numerical integration
and wave-power iteration for all 25 ions at 1.069 $R_{\odot}$.
Recall that {\em no assumptions} were made about the frequency dependence
of the resonant wave power spectrum for these models.
The power for each ion was computed independently of the others.
The resonant wave power is shown as a function of $Z_{i}/A_{i}$ for
each of the 16 choices for the $T_p$ normalization.
Straight-line fits are shown for two distinct regions of
$Z_{i}/A_{i}$ values, in order to guide the eye and make rough
comparisons to the theoretical expectations for ${\cal P}(\omega)$.
The black curves in Figure~\ref{steve4} are shown to compare the
semi-empirical results with the expectations of turbulence theory.
In other words, these curves come from evaluating
equation (\ref{eq:Pomega}) numerically for ${\cal P}(\omega)$,
assuming a range of $\beta / \gamma$ values between 0.01 and 1.

For large enough choices for $T_p$, there occur several ions that
have smaller perpendicular temperatures than the adopted proton
temperature at 1.069 $R_{\odot}$.
These are shown with dashed lines for the straight-line fits, and
thus should not be considered as self-consistent models for the
entire ion data set.
In Figure~\ref{steve4}, the ``green'' curves indicate a
collection of models all
with $T_{p} \approx 2.1$ MK, because we wanted to explore specifically
the behavior of the O$^{5+}$ heating in the limit of strong
collisionality (i.e., because $T_{\perp i} \approx 2.15$ MK for this
ion); see below.

For the lowest frequencies (i.e., $Z_{i}/A_{i} \lesssim 0.27$),
both the spectral slope and the absolute wave power at 1.069 $R_{\odot}$
seem to agree well with the turbulent cascade predictions for
values of $\beta / \gamma \lesssim 0.3$.
This agreement is actually quite notable, since the theoretical
predictions (black curves) were normalized by pre-existing assumptions
about, e.g., the total wave power $U_{\rm tot}$, and the empirical sets
of points were iterated completely independently from any considerations
of MHD turbulence.
If the proton temperature in the low corona remains less than about
2 MK (see, e.g., Figure~\ref{steve2}), then this result provides some
additional observational justification for a value of $\beta / \gamma$
of order 0.1 to 0.3 (see Cranmer \& van Ballegooijen 2003).

For the highest frequencies ($Z_{i}/A_{i} \gtrsim 0.27$), it seems
clear that some kind of ``upswing'' in the resonant wave power is
indicated by the empirical points in Figure~\ref{steve4}.
If this is indeed a real effect, there are several possible explanations
in terms of MHD and kinetic fluctuations:
\begin{enumerate}
\item
The additional power at large gyrofrequencies may arise because of
{\em plasma instabilities} centered around the cyclotron resonances of
alpha particles ($Z_{i}/A_{i} = 0.5$) or protons ($Z_{i}/A_{i} = 1$);
see, e.g., Markovskii (2001), Isenberg (2001), Zhang (2003),
Laming (2004), and Markovskii et al.\  (2006).
Thus, the effects seen at charge-to-mass ratios around 0.35 may
just be the tail of an additional population of waves that peak at
larger frequencies.
\item
This may be a similar effect as the spectral flattening (i.e.,
enhancement above a power law frequency dependence) evident in radio
scintillation measurements of density fluctuations at larger heights
(e.g., Coles \& Harmon 1989).
This flattening was modeled successfully by Harmon \& Coles (2005)
by taking into account the enhanced compressibility of {\em obliquely
propagating Alfv\'{e}n waves} once they reach ion cyclotron frequencies.
It may be possible for compressibility effects to either alter some
aspects of the turbulent cascade or change the functional dependence
of the heating rate on the available wave power (i.e.,
eq.~[\ref{eq:Qzone}]) in order to produce this effect.
\item
A somewhat more speculative idea is that the particular type of
MHD turbulence acting in the low corona may undergo a kind of
{\em bottleneck effect} where the wave power piles up near the
dissipation range.
This effect appears in many turbulence simulations, and it has
often been suspected of being numerical in origin (e.g.,
Verma \& Donzis 2007).
However, there have been independent theoretical predictions of similar
kinds of wave-power pileup arising from nonlocal interactions between
disparate scales in $k$-space (e.g., Falkovich 1994;
Biskamp et al.\  1998).
\end{enumerate}

\begin{figure}
\epsscale{1.13}
\plotone{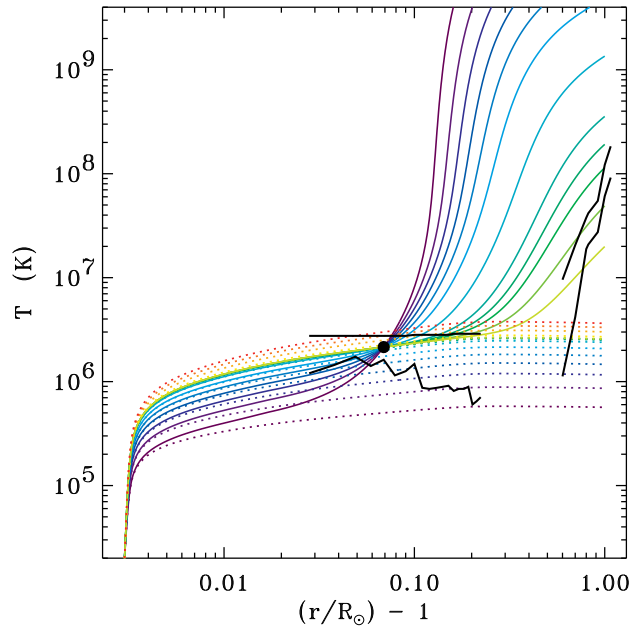}
\caption{Radial dependence of modeled proton temperatures
({\em dotted lines}) and perpendicular O$^{5+}$ temperatures
({\em solid lines}), with the same color key as Figure~\ref{steve4}.
Also shown are observational lower and upper limits on $T_{\perp i}$
from SUMER and UVCS ({\em black solid lines}).}
\label{steve5}
\end{figure}

Figure~\ref{steve5} illustrates the radial dependence of the ion heating for
O$^{5+}$ ions, and shows how $T_{\perp i}$ varies with the different
choices for $T_{p}$.
The color scheme for proton temperature values is the same as in
Figure~\ref{steve4}.
The black curves at the heights corresponding
to the SUMER measurements show the lower and upper limits defined by
equation (\ref{fwhm_4}).
The black curves at larger heights give the limits on the ion
temperature from recent UVCS \ion[O vi] empirical models
(Cranmer et al.\  2008).

The models with relatively low proton temperatures ($T_{p} \lesssim 2$ MK)
require a large amount of resonant wave power to ``combat''
Coulomb collisions and raise the ion temperature to the observed
SUMER value at 1.069 $R_{\odot}$.
Unfortunately, this large resonant wave power then causes the
ion temperature to ``explode'' to unrealistically huge values at
larger heights where the plasma becomes collisionless.
On the other hand, the green curves in Figure~\ref{steve5} agree reasonably
well with both the SUMER and UVCS data and correspond to larger
assumed values for $T_p$.
Here, the reason for a fine grid of proton temperature curves having
\begin{equation}
  0 \, < \, \frac{T_{\perp i}}{T_p} - 1 \, \ll \, 1 
\end{equation}
becomes evident.
These choices for $T_p$ do not need as much wave power to reach
the observed SUMER ion temperature at 1.069 $R_{\odot}$, and thus
they are less likely to result in unrealistically large temperatures
at larger heights.
However, it is important to note that these best-fitting solutions
seem to have quite strong Coulomb collisions even at the {\em lowest}
of the SUMER heights.\footnote{%
It is possible, of course, that Coulomb collisions are not the only
means of thermalizing and isotropizing the ions.
The existence of various kinds of turbulent fluctuations has been
suggested to be able to can couple particles and possibly provide
extra ``quasi-collisions'' (e.g., Perkins 1973; Dum 1983; Kellogg 2000).
This may need to be investigated in the context of the low corona.}
This would imply that at, say, 1.03--1.05 $R_{\odot}$, all of the ion 
temperatures would be expected to be tightly coupled to both $T_p$ 
and to one another. This, as seen in Figure~\ref{res1} above, is not 
the case. Clearly the models presented here are only a first step 
toward a self-consistent description of the preferential ion heating 
that applies for the full range of heights observed by SUMER and UVCS.

\section{Conclusions}
\label{conclusions}

In the present work we have used SUMER observations of a polar coronal hole
to measure the ion temperatures $T_i$ for a large number of ions, in order
to determine their dependence on the charge-over-mass ratio $Z_{i}/A_{i}$.
We repeated our measurements for several heights between 1.03 $R_{\odot}$
and 1.17 $R_{\odot}$
in order to investigate the height dependence of our results. We used the 
method devised by Tu \etal (1998) to determine $T_i$; this method only assumed
that non-thermal velocities $v_{nth}$ are the same for all ions. 

We found that ion temperatures are larger than the electron temperatures in 
nearly all cases and at all heights, and that results are qualitatively the 
same in the 1.03--1.17 $R_{\odot}$ range of distances from the limb, although
the number of available ions decreases as distance from the limb increases. 
Our most notable result is an U-shaped dependence of $T_i$ from
$Z_{i}/A_{i}$, where ions with low $Z_{i}/A_{i}$ (less than about 0.23)
have very large $T_i$ values, while
those with $Z_{i}/A_{i} \gtrsim 0.23$ have roughly constant $T_i$.

We used our measured $T_i$ values to constrain an exploratory model
of solar wind heating and acceleration based on ion-cyclotron waves,
under the assumption that such waves are gradually replenished as they
are dissipated. The measured $T_i$ values
help us constrain the proton temperatures as well as the ratio
between advection and diffusion (in a likely scenario of anisotropic
MHD turbulence).
We find that the observations of ions having $Z_{i}/A_{i}$ values smaller
than about 0.25 are consistent with a turbulence model very similar to
that shown by Cranmer \& van Ballegooijen (2003) to be able to energize
protons in the extended corona.

However, $T_i$ measurements obtained for ions with $Z_{i}/A_{i}$
larger than 0.25
show an upswing in wave power that is difficult to reconcile with
traditional views of turbulent cascade.
We discuss the implications of this inferred increase in wave power,
suggesting several different possibilities such as plasma 
instabilities, obliquely propagating Alfv\'en-wave compressibility
effects, or a turbulent bottleneck effect.

\acknowledgements
The work of Enrico Landi is supported by the
NNG06EA14I, NNH06CD24C as well as other NASA grants.
The work of Steven Cranmer is supported by NASA under grants
{NNX\-06\-AG95G} and {NNG\-04\-GE77G} to the Smithsonian
Astrophysical Observatory.
The authors would like to thank Adriaan van Ballegooijen and
Mari Paz Miralles for valuable discussions.

\end{document}